\documentclass[11pt]{article}
\usepackage{graphicx}
\usepackage{amssymb,amsmath,amsfonts,palatino,amsthm}
\usepackage{amssymb}
\usepackage{epstopdf}
\newcommand{\beq}{\begin{equation}}
\newcommand{\eeq}{\end{equation}}

\newcommand{\f}{\begin{equation}}
\newcommand{\ff}{\end{equation}}

\begin{document}

\title{On limitations of the extent of inertial frames in non-commutative relativistic spacetimes \\}
\author{Lee Smolin\thanks{lsmolin@perimeterinstitute.ca} 
\\
\\
Perimeter Institute for Theoretical Physics,\\
31 Caroline Street North, Waterloo, Ontario N2J 2Y5, Canada}
\date{\today}
\maketitle
\vfill

\begin{abstract}

We study the interplay of non-locality and lorentz invariance in a version of deformed or doubly special relativity (DSR) based on kappa-Minkowski spacetime. We find that Einstein's procedure for an inertial observer to assign coordinates to distant events becomes ambiguous for sufficiently distant events. The accuracy to which two clocks can be synchronized turns out to depend on the distance between them. These are consequences of the non-commutativity of space and time coordinates or a dependence of the speed of light on energy in relativistic theories. 

These ambiguities grow with distance and only become relevant for real observations for the description of cosmologically distant events.  They do not afflict the interpretation of the detection of gamma rays in stationary or moving frames near the detector. Consequently there is no disagreement between the principles of DSR and the observation that interactions in nature are local down to currently observable scales.

\end{abstract}
\vfill
\newpage
\tableofcontents

\section{Introduction}

Deformed or doubly special relativity is an hypothesis about how the principle of the relativity of inertial frames can be made consistent with the existence of a minimal
length scale, taken to be the Planck length, $l_{P}$\cite{DSRI,DSRII,DSR-reviews}.   It may also be seen as making possible a maximum momenta for an individual particle.  Over the almost decade since it was first proposed
DSR has been realized in a number of frameworks.  The most developed of these are in $2+1$ dimensions\cite{2+1}, and these give confidence that the idea can be sensibly realized in the context of a quantum field theory.  At the same time, there is not yet a completely developed realization in $3+1$ dimensions.   

Nonetheless there is an expectation that at least some versions of the idea combine the relativity of inertial frames with an energy dependence of the speed of light.  
To leading order in $l_P$ this would have the form
\f
v(E) = c(1 -\alpha l_P E/\hbar ) 
\label{c(E)}
\ff
for a dimensionless parameter $\alpha$. 
If so, this has implications for observational tests of lorentz invariance at linear order in the Planck length\cite{GL-phenom}.

A major issue in the interpretation of $DSR$ theories has been the presence of non-local effects, at least at the Planck scale.  
This has been discussed by a number of authors\cite{sabine-paradox,sabine-preprint,Unruh-DSR,GAC1,JKG1,ALN, Rychkov,nonlocality,GTetal}.  
So far it has been unclear whether these non-local effects destroy the consistency of the theory or exactly what the correct physical interpretation of these non-local effects are.  This non-locality is tied up with the realization of lorentz invariance in $DSR$ theories.  At issue is whether interactions which are local in one frame of
reference, become non-local when described in the coordinates defined by frames moving relative to them. A serious issue raised in \cite{sabine-paradox,Unruh-DSR} is the possibility that the non-local effects generated by lorentz transformations are non-local, in a way that 
lead to manifest conflict with the body of experimental evidence that supports the postulate that physical interactions are local.  
An important question to resolve is whether these non-localities compromise the interpretation of experiments underway in which 
(\ref{c(E)}) is tested, such as in observations of gamma ray bursts by the Fermi satellite\cite{GL-phenom}.

The present paper has two aims.  The first is to study the description of processes in which gamma rays are emitted and detected, as it appears 
in frames of reference moving with respect to the detectors.  We carry this out in a well studied 
approach to $DSR$, which is $\kappa-$Minkowski spacetime\cite{KP1,KP2,KM1,KM2,previous-free,JM1,JM2}.  
We find that while there are no macroscopic non-localities seen by observers close to the detecor, there are issues with ambiguities in coordinates of distant events. 

The second aim of this paper is to propose a new feature of theories which incorporate deformations of special relativity, which is a limitation on the spatial extent
to which an inertial frame can be defined, coming from ambiguities in the procedure for synchronizing clocks which arise due to the non-commutativity of the spacetime geometry or to the energy dependence of the speed of light (\ref{c(E)}).  In special relativity the coordinates of an inertial frame
are defined in terms of a single clock at the origin, and exchanges of light signals are used to define coordinates for distant events.  As we show here,
this procedure breaks down in $\kappa$-Minwkowski spacetime, so that ambiguities can appear in the coordinates assigned to distant events.   
This novel effect is a kind of uv/ir mixing, applied to the operational definition of spacetime coordinates.  As a consequence, the ambiguities in the coordinates
of distant events are not signals of real physical non-localities, and there is no conflict with experimental evidence for locality on macroscopic scales.  

These conclusions differ from those of \cite{sabine-paradox}, which however works in a different framework, based on different assumptions.  The reasons
for, and consequences of, this disagreement are explored in an Appendix\footnote{In a note added below we comment on the recent critique of this paper \cite{sabine-critique}.}. 

In the next section we discuss the description of a detector of gamma rays arriving from a distant burst, from the point of view of an observer moving relative to the detector, in $\kappa$-Minkoski spacetime.  In section 3 we discuss the attribution of coordinates to events far from the origin of
a lorentz frame and propose that ambiguities in the coordinates are to be understood as coordinate ambiguities rather than physical non-localities. In section 4 this view is supported by an analysis of Einstein's procedure for the synchronization of moving clocks, applied to  $\kappa$-Minkowksi spacetime. This 
section closes with some general remarks about the physical interpretation of relativistic but non-commuting spacetimes.   
In the conclusion we comment on  the present status of the issues discussed here.

\section{Gamma ray bursts and moving observers}  

The focus of this paper is an experiment in which photons from gamma ray bursts are observed to arrive near Earth. Such experiments are of interest because they have already been used to test the hypothesis of an energy dependence of the speed of light, eq (\ref{c(E)}) \cite{GL-phenom}.   We will be interested in how the experiment is described in terms of the coordinates constructed by different inertial frames, in order to understand the implications for it of the
hypothesis that $DSR$ preserves the relativity of inertial frames.  

We first describe the experiment 
in an inertial reference frame which can be assumed to be at rest on the Earth and to have its
origin situated at a detector on the Earth.  We will call this Alice's frame of reference. 

 Long ago and far away there was a gamma ray burst.  In it there was an event, $\cal E$, where an atom
emitted two gamma rays, $\gamma_1$ and $\gamma_2$, both moving in the positive $\hat{x}$ direction, which we take to be oriented towards the Earth, with energies $E_1$ and $E_2$.  We will assume
that $E_2 >> E_1$.  The coordinates of the event $\cal E$ in Alice's frame are ${\cal E}^a =(t, x)=  (-L/c, -L)$. 
We assume that the speed of a photon is given by (\ref{c(E)}).

Alice has a photon detector at the origin of her coordinates.  It consists of a device which amplifies the effect of a photon scattering from an electron.  It then contains 
some electrons, which are passing though the detector, and can be considered to be moving slowly.  At a time $t=0$, on her clock, the first photon encounters an electron in her detector.  This detection event,
${\cal F}_1$ then occurs at coordinates ${\cal F}_1^a= (0,0)$ in her frame.  At a later time $\tau$ the second photon arrives and encounters an electron which is just passing through the detector at that time, which is also amplified.  This is the second detection event ${\cal F}_2$ which occurs at coordinates  ${\cal F}_2^a= (\tau,0)$ in her frame\footnote{To facilitate comparison with \cite{sabine-paradox}, we note that these coordinates are shifted slightly from those used there; in that reference ${\cal F}_2^a= (0,0)$ 
while ${\cal F}_1^a= (-\tau,0)$.  As the shift is already of order $t_p L $, this does not affect the conclusions of the analysis to leading order in $t_p L $. }.

It is easy to compute that
\f
\tau = \frac{\alpha t _P \Delta E L}{\hbar c^2}
\label{timedelay}
\ff
where $\Delta E= E_2 - E_1 $.

We now consider how this experiment is described in a frame of an observer, called Bob, who is in a satellite which passes by Alice with a velocity $v$ at the time $t=0$. Let us call this Bob's frame of reference\footnote{In \cite{sabine-paradox} this is called the satellite frame.}. 
 We then 
put the origin of Bob's frame so that the event ${\cal F}_1^a= (0,0)$ in Alice's frame has the coordinates ${\cal F}_1^{a \prime} = (0,0)$ in Bob's frame.
This implies that the coordinates of events in Bob's frame are found by making a passive lorentz transformation at the event ${\cal F}_1$. 

\subsection{Description of the experiment in the $\kappa$-Poincare framework}

To define the lorentz transformation to Bob's frame we work in $\kappa$-Minkowski spacetime\cite{KM1,KM2,previous-free,JM1,JM2}.
In this framework the Lie algebra of the Poincare group is deformed to a quantum algebra called the $\kappa$-Poincare algebra.   
This is to allow the time and space coordinate to fail to commute,
\f
[x^i , t ] =\imath t_p x^i
\label{[x,t]}
\ff
This can be further seen as a consequence of a postulate the momentum space is curved.

A feature of $\kappa$-Minkowski spacetime is that the $\kappa$-Poincare algebra is non-linear and has a non-linear action on 
coordinates.  Consequently, one has some freedom in defining what quantities in the phase space  of a free relativistic particle correspond
to physical spacetime coordinates, physical  momentum and physical energy, as would be measured by macroscopic detectors.  
Some argue that physics should be invariant under these choices, while others argue that a single choice is correct.  Because this issue is not resolved
we will study here two hypotheses.  The first is that particles propagate along worldlines defined in the non-commuting coordinates
$(x^i, t)$.   It has been argued that worldlines defined in these coordinates have an energy-{\it independent} speed of 
light\cite{previous-free,JM1,JM2}.  Nonetheless we will see that there are issues with
apparent non-localities or coordinate ambiguities.  

The second hypothesis is that physical particles propagate along worldlines defined in a commuting set of coordinates, which differ
from the first by taking physical time to be measured by\cite{previous-free,me-firstresponse}
\f
T=t+\frac{t_p}{\hbar}x^i p_i e^{-t_p E}
\ff
$T$ satisfies 
\f
[T,x^i]=0
\ff
This choice does lead to an energy dependent speed of light\cite{previous-free,me-firstresponse}.

We consider each in turn.

\subsubsection{The moving frame in non-commutative coordinates}

We construct the measurements made in Bob's frame of reference by doing the explicit lorentz transformation.

It is important to note that in $\kappa$-Poincare the effect of a boost is dependent on the energy of the particle whose trajectory is being boosted. 
The formulas for carrying out a lorentz transformation in $\kappa$-Poincare are given in 
 \cite{JM1,JM2}.   A pure boost  denoted by a
spatial vector $\omega^i$ is given by\footnote{Here and in the following we set $\alpha =1, t_p = l_p/c$ and $E_p = \hbar c/ l_p$.  I am inconsistent about
factors of $\hbar$ and $c$.}
\f
\delta x^i = - \omega^i t - \frac{1}{E_p} \epsilon^{ijk}\omega_j L_k
\label{xtrans}
\ff
\f
\delta t = -\omega \cdot x + t_p \omega \cdot N
\label{ttrans}
\ff
where $L_i$ are the spatial angular momentum generators
\f
L^i =  \epsilon^{ijk} x_j p_k e^{-t_p E}
\ff
and $N_i$ are the generators of deformed boost transformations
\f
N_i =- p_i e^{-\frac{p_0}{E_p}} t - x_i \left [ \frac{E_p}{2} (1- e^{-\frac{2p_0}{E_p}} ) + \frac{1}{2E_p}p\cdot p e^{-\frac{2p_0}{E_p}} \right ]
\label{boosts}
\ff
Note that to leading order in $t_p$ (\ref{ttrans}) implies
\f
\delta t = -\omega \cdot x [1+ t_p E] + t_p \omega \cdot pt
\label{ttrans2}
\ff

Let us consider how the three events in the experiment look in Bob's frame.  
To get to the description in Bob's frame we make a pure boost, using (\ref{xtrans}) and (\ref{ttrans}).  We note that, just as in special relativity, it is 
completely unambiguous what Lorentz transformation to use; this is fixed by the requirement that the origin of Bob's coordinates coincide
with the origin of Alice's coordinates. This means that the event  ${\cal F}_1^a= (0,0)$ is fixed by the pure boost\footnote{The results don't change
to leading order in $t_p L$  if instead ${\cal F}_2^a= (\tau,0)$ is taken to be the fixed point. See Footnote 1.}. 
However, some of the results depend on the energy, and hence one has to choose which particle's energy, among those whose coincidence
defines the event, is used to define the coordinates in the boosted frame.  

We find from (\ref{xtrans}) and (\ref{ttrans}) that the  first detection event has unique coordinates in Bob's frame, because it corresponds to the simultaneous origin of both the coordinates of Alice's and Bob's frame.  

We next find that the second detection event is  split, in that the event is given two coordinates, depending on which particle is used for the
transformation.  But this is a very tiny effect as it occurs  only at order $t_p^2$.  Explicitly, ${\cal F}_2^{0 \prime}$, which is the
time coordinate of the second detection event in Bob's frame, is given two values, separated by a time difference,
\f
\Delta \tau^\prime = \frac{ v t_p^2 E_2^2 L}{\hbar^2 c^2}
\label{realresult}
\ff
We see that this  split must be
proportional to $\tau t_p$, which is of order $t_p^2$.  For the most energetic gamma rays detected by the Fermi satellite, in which 
$E$ is on the order of $10 Gev$, $v \approx 10^{-5}$  and $L$ is cosmological, this is of the order of $10^{-24}$ sec, which is not detectible in the experiment.  
 
We note that there is one factor of $L$ which is in $\tau$ and comes from the fact that the delay is the result of a long travel time.  But to get an observable effect there would have to be another factor of $L$ coming from the boost, to balance the $t_p$ in the energy dependent
part of the boost.  But this is impossible as the boost is defined just at the detector and is being applied to a nearby event.    
How can it matter for a lorentz boost of the 
trajectory of a photon near the origin of the reference frame how long that photon has been traveling?

Now let us look at the emission event.  
The emission event is for Bob split into two events ${\cal E}^{ \prime}_1$ and  ${\cal E}^{ \prime}_2$, corresponding to the creation of the two photons
$\gamma_1$ and $\gamma_2$.  If we assume that both photons have the same angular momentum their space  coordinates are the same
${\cal E}^{ \hat{x} \prime}_1 = {\cal E}^{  \hat{x}  \prime}_2$.  However there is a split in their time coordinates given by 
\f
\Delta t_{{\cal E}^\prime} = - t_p \left ( \omega \cdot \vec{x} \ \Delta E + \omega \cdot \vec{\Delta p} \ t   \right )
\label{effect1}
\ff
where $\omega_x$ is the component of the boost in the $\hat{x}$ direction.

Because $L$ is a cosmological distance, this can be a macroscopic time interval.  
Thus, Bob sees that the second photon is emitted at a time later or earlier than the second photon, depending onthe direction of Bob's motion.    In fact, in the case under discussion we have 
\f
\Delta t_{{\cal E}^\prime} = t_p \omega_x L  ( \Delta E + \Delta p )
\label{effect2}
\ff
Note that to  
leading order in $t_p$,  $E \approx |p| $ so that, neglecting the low energy photons momentum,  and taking into account that the
photon is moving in the positive $\hat{x}$ direction, so $\Delta p >0$, we have $\Delta E \approx \Delta p $
Thus, 
\f
\Delta t_{{\cal E}^\prime} =2  t_p \omega_x L  \Delta E 
\label{effect3}
\ff
It is interesting to note that in the case that the photon is moving away from the Earth, we would have
\f
\Delta t_{{\cal E}^\prime}^{\mbox{photon moves away}} = O  (t_p^2)
\label{effect4}
\ff

\subsubsection{The moving frame in commuting coordinates}

We now consider how the experiment looks in commuting coordinates, $(T,x^i)$\cite{previous-free,me-firstresponse}.
We can mention that one reason to suppose this is the physical choice is that there is a basis in the Hilbert space
for a free relativistic particle in which $x^i$ and $T$ are both sharply defined, while is there is no basis
that simultaneously diagonalizes $t$ and $x^i$, because they don't commute.  So if the classical physics arises
as a limit of quantum physics, this is more likely to happen in the case of commuting coordinates. 

We note that $T$ translates and lorentz  boosts conventionally, so that under boosts
\f
\delta T = -\omega \cdot x
\ff
Meanwhile the transformation law for $x^i$ under lorentz boosts is modified to
\f
\delta x^i = \omega^i (T- t_p x \cdot p ) + O(t_p^2) 
\ff

We ask how the three events in the gamma ray production and detection appear in Alice and Bob's coordinates when these
$(T,x^i)$ coordinates are used.  A simple calculation shows that both detection events are unsplit in both frames.  However we encounter
a surprise when examining the emission event, as it is split even for Alice when she uses the time coordinate $T$.  Assuming that the two 
photons were emitted at the same time $t$ in Alice's frame, the commuting time, $T$, that
each gamma ray was emitted is shifted to
\f
T= -L (1 + t_p p )
\ff
so that the emission event is split by a time interval 
\f
\delta T_{emission, Alice} = - t_p L \Delta p /\hbar
\label{Tsplit-alice}
\ff
It turns out that the lorentz transformation doesn't change this, so that also
\f
\delta T_{emission, Bob} = - t_p L \Delta p/\hbar
\label{Tsplit-bob}
\ff
Finally, both Alice and Bob see the spatial coordinate unsplit.  

Of course, one could instead proscribe that the two photons are emitted at the same initial value of $T$, in which case there is no splitting in Alice's frame, 
but there is a splitting in Bob's frame of the spatial coordinate of the emission event.  
\f
\delta x_{Bob}^i= t_p \omega^i L \Delta p
\label{effectx}
\ff

\section{Coordinate ambiguities for distant events}

As we have just seen, and as was discussed also in \cite{me-firstresponse},  a boost appears to split the time coordinate, $t$, of events distant from the fixed point of the boost, because a single event involving the coincidence of two or more particles can be given several distinct time coordinates depending on the energy of each particle.  However these
energy dependent effects in boosts are proportional to $t_p x E/\hbar$ or $t_p t p /\hbar$,  so they are only sizable for events far from the origin around which the boost is defined.  The question is whether these splittings in coordinates of events far from the origin of a boost, of order $l_P x E/\hbar$ times the boost parameter $\beta$ are real physical non-localities or coordinate artifacts.  

In addition, we have seen that if we use the commuting time coordinate, $T$ that both Alice and Bob see the emission event to be split
by (\ref{Tsplit-alice}).  This would certainly seem a coordinate artifact as it is introduced by rescaling the time coordinate by a term which 
is momentum dependent.  

It is also interesting to note that in the $\kappa$-Poincare context translations can also split the non-commuting time coordinate, $t$, of an event.  The formula for an 
infinitesimal translation in $\kappa$-Poincare is\cite{JM1,JM2}
\f
\delta t= a^0 - t_p p_i a^i e^{-t_p p_0 } ; \ \ \ \ \ \ \ \delta x^i = a^i
\label{translate}
\ff
for infinitesimal translations labeled by a four vector, $(a^0, a^i)$.  Thus, consider an event at $t=0$, $x^i=0$ in some frame, defined by
the collision of two particles with momentum, $p_1^i$ and $p_2^i$.  If we consider this event with respect to another frame at rest with
respect to the first, but translated a distance $a$ in the $\hat{x}$ direction, the event will be split into two events with a time
difference
\f
\Delta t^\prime = t_p a (p^x_1 - p^x_2 ) + O(t_p^2 )
\ff

So also we have to ask if what we have here is a case of one event that somehow has become two, or simply a single event whose
time coordinate in a distant frame is ambiguous.  I would claim that  for both for boosts and translations, these  coordinate ambiguities are just a new kind of coordinate artifact.   To support this I would note that these coordinate ambiguities have very peculiar properties for real physical effects.  

\begin{itemize}

\item{} The alleged problem with locality always occurs at very large distances from the point around which boosts are made, ie the origin of the coordinate system that defines an inertial frame.  

\item{}The  presence or absence of this distant non-locality appears to depend on the position and motion of the observer.  What happens is that an event which is local in one reference frame,
appears to become two events, when described with the coordinates of a frame of reference which is moving with respect to the first and/or very distant from it.  

\item{}When a single event is split by such a distant lorentz boost into two, they are time like separated, and which is to the future and which past depends on the direction of the
velocity of the moving frame.   

\item{}When the event is split by a translation, the causal order of the two events that are apparently created depend on whether the translations was done to the left or to the right.   When the event is split by the use of a commuting time coordinate, $T$, the causal order depends
on the direction of the momenta of the photons.

\end{itemize}

I would then propose that the correct interpretation of these results is that the apparent split, in which the time coordinate of a single distant event is given two or more different values, is not a physical phenomena at all, but only a coordinate ambiguity, which occurs when one attempts to define a moving inertial frame by synchronizing clocks over large distances.  How else could we describe the fact that whether the event has one or two time coordinates depends on which reference frame is being used to describe the event? And how else can we understand the
fact that  the choice of which of two apparently time like separated events is to the future of the other depends on the direction of motion of a very distant observer? 

This is classical physics, so whether an event is to the causal future or past of another event cannot be influenced by the direction that may be chosen for boosting an observer who is not only spacelike from the events in question, but at a cosmological distance from it.  Therefor causality requires that the alleged splitting of very distant events into two is an artifact of a coordinitization procedure that has become ambiguous.

I will support this in the next section, where I  will discuss how a coordinate ambiguity arises from the attempt to apply Einstein's procedure for synchronizing clocks to the case of $\kappa-$ Minkowski spacetime.  The result is that, unlike Minkowski spacetime, inertial frames cannot be extended indefinitely without running into coordinate ambiguities.  
 
\section{Synchronization of distant clocks}

Let us recall  Einstein's operational construction of coordinates for an inertial observer's measurements\cite{Albert}.  According to Einstein,
events are {\it defined} by the physical coincidence of elementary particles, which in classical physics means by the intersection of the worldlines of those particles. 
Coordinates are {\it assigned to those events}  by physical operations which involve exchanges of light signals between those
events and the observer at the origin, who is also presumed to carry a clock.  They have no other meaning.  It is assumed that the observer
has no access to distant events which would enable her to assign them coordinates except by the exchange of light signals. 

Thus, in Einstein's procedure,  you start with an observer, Alice, who has a clock next to her.  Events on the worldline
of the observer are parameterized by the reading of the clock, $t$.  Events off the worldline are given time and spatial coordinates
by using light signals bounced between the observer and those events.  The only measurements that are made are of the readings
of the clock when the light signals leave and return, $t_1$ and $t_2$, having bounced back off of the distant event.    
An event, $e$ is assigned a value of the time coordinate, $t_e = (t_1 + t_2 )/2$.  The event is assigned a space coordinate (simplifying to
$1+1$ dimensions) of $x_e = (t_2 - t_1 )c/2$.   A consequence of this is that the observer is by definition at the event $x=0$. 

A by product of this construction is that it allows the synchronization of distant clocks with the observer's clock.  Another consequence is that,
when applied to moving clocks, this leads to the relativity of simultaneity.  

In special relativity this procedure can be applied to events arbitrarily far from the observer. In general relativity there is a limitation to
how large $x$ and $t$ can be before the procedure becomes ambiguous, which is given by the
radius of curvature. The curvature limits the region of space and time over which clocks can be synchronized.  

The results we have discussed raise the question of whether in DSR, or quantum gravity generally, this procedure can be unambiguously
extended to arbitrarily large $x$ and $t$, or whether there are quantum effects, even in the absence of classical curvature, which limit
the applicability of this procedure to arbitrarily large values of the coordinates.

Here I would like to point out that there is in fact a limit to how large of a region of space and time may be unambiguously be
assigned coordinates in an inertial frame.  This limit arises because  any clock has an accuracy within which one like to synchronize it.  We will call this
$\omega_0$.    

To see how this introduces a limit to how far away two clocks can be and be synchronized with each other, let us recall 
from \cite{me-firstresponse} that a massless particle propagating in $\kappa$-Minkowski spacetime (in the commuting coordinates) 
is subject to an anomolous
spreading of the wavepacket.  An initially Gaussian wavepacket, with initial width $\Delta x$ has a width that
evolves according to (eq. 61 of  \cite{me-firstresponse})\footnote{An equivalent result was found in eq. 9 of  \cite{sabine-paradox}.}
\f
\Delta w = \sqrt {\Delta x^2 +  x^2 l_p^2 \Delta p^2/\hbar^2 }
\ff
where $\Delta p = \hbar/\Delta x$.  The minimal width after traveling a distance $x$ is then
\f
\Delta x_{min} =  \sqrt{2 l_p x }
\ff
This implies a minimal uncertainty in the arrival time, $t_2$ of a photon exchanged with a distant event and hence a minimal uncertainty
in the time $t_e$ attributed to that event of 
\f
\Delta t_{e} = \frac{1 }{c} \sqrt{2 l_p x }
\ff
If we want this uncertainty to be less than the inverse frequency $\omega_0^{-1}$, within which we demand the synchronization is accurate, we
find,
\f
x < \frac{c^2}{2 l_p \omega_0^2}
\label{horizon2}
\ff
The same result can be derived for the non-commuting coordinates, which satisfy (\ref{[x,t]}).  
This implies the uncertainty relation
\f
\Delta x \Delta t \geq t_p |x|   .
\label{nup}
\ff

Let us consider that we are trying to synchronize a light clock, constructed by bouncing light between mirrors a distance $\Delta x$ apart,
placed a distance $| x|$ away from the origin of a coordinate frame.
The clock will have a frequency $\omega_0 = c/\Delta x$, and it will allow time to measured to an accuracy 
$\Delta t \approx \omega^{-1}_0$.   So we have
\f
\Delta x \Delta t  \approx  \frac{c}{\omega^2_0}  .
\label{nup2}
\ff
which implies $\frac{c}{\omega^2_0} > t_p |x|$.  
From this we conclude that such a clock measures time in a way that is inconsistent with 
the uncertainty relation (\ref{nup}) unless it is within a distance from the origin $|x|$ bounded by (\ref{horizon2}).  That is, any quantum
dynamics which respects (\ref{[x,t]}) and (\ref{nup}) must make it impossible to synchronize a clock distant from the reference clocks
that defines the origin of a reference frame, unless (\ref{horizon2}) is respected.

These general limitations (\ref{horizon2}) and (\ref{nup2}) are strengthened when the distant event to which  the observer wishes to assign coordinates
involves an energetic photon.  This is shown by the following two considerations. 

First, note that the relativity of inertial frames means it can't matter which frame one starts with, all frames are equivalent. 
Hence the procedure of assigning inertial coordinates to distant events can only be as accurate as allowed by
the ambiguities produced by the lorentz transformations.  So we can check that the same limitations arise from requiring 
that the energy dependent terms in the lorentz transformations not affect the synchronization of clocks.

Suppose the observers want to assign a time coordinate to an event involving several photons with a range of energies $\Delta E$, using clocks that are reliable within an  accuracy of $\omega_0$   This means that we must have $\Delta t < \omega_0^{-1}$
which implies that 
\f
L <R(\omega_0, \Delta E) =  \frac{\hbar c}{t_p\omega_0 \Delta E}
\label{horizon1}
\ff
The radius $R(\omega_0, \Delta E)$ is a limit of how far an event can be from the reference clock and still be given an unambiguous time coordinate, within an accuracy of $\omega_0$, when that event involves
photons in an energy range $\Delta E$.     Now, if we are using photons to synchronizing clocks we are free to use photons of any energy so long as 
$E > \hbar \omega_0$.   So given only the accuracy of the synchronization we want, we find again (\ref{horizon2}).  

An independent argument for (\ref{horizon1}) comes from considering an experiment where the observer has synchronized a network of clocks with an accuracy of $\omega_0$ at rest with respect to her.  The observer wants to assign a space and time coordinate to an event, $P$, a distance, $L$ from them, which  involves the detection at that point of a photon with a large energy up to $\cal E$.  To do this she sends a photon with energy $E_1$ to a the clock at $L$ which immediately returns a photon of energy $E_2$.   To avoid a net momentum transfer which would move the clock, disrupting its synchronization and use in an inertial system of coordinates,  the clock passes on the momentum of the  incident photon so  $E_2 \approx {\cal E}$, which means $\Delta E > {\cal E}$.  In this case the uncertainty of the time at which the second photon is $\delta t_2 = \sqrt{l_p L {\cal E}}$.  This gives 
 \f
L <R(\omega_0, \Delta E) =  \frac{\hbar c}{t_p\omega_0 {\cal E}}
\label{horizon3}
\ff
Thus, the stronger bound (\ref{horizon3}) is necessary if the observation of the energetic photon by the distant clock is not to disrupt the synchronization, rendering the system useless for making further reliable measurements of the inertial coordinates $x$ and $t$.  

All four arguments lead to the conclusion that there is a limit to how far apart two clocks can be if we want to synchronize them to a given
accuracy, $\omega$.    The first two arguments establish that there is a limit, which arises from the need to use photons of finite energy
to synchronize distant clocks to a given precistion.  However if the event that one is using exchanges of light signals to coordinatize involves a very
energetic photon, there is a tighter limit, given by  (\ref{horizon1}) and (\ref{horizon3}).  These are necessary to keep the distant clock synchronized
after the absorption of the energetic photon. 
These limits (\ref{horizon1}) and (\ref{horizon3}) are sufficient to explain the apparent non-localities discussed
earlier as coordinate ambiguities. 

This phenomena is a kind of infrared/ultraviolet mixing, in that the presence of non-commutivity or modified dispersion relations at the Planck scale
are limiting the definition of inertial coordinates at a large scale.  This accords with the
point of view proposed in \cite{me-firstresponse} that when $l_p = \sqrt{\hbar G/ c^3}$ is present, quantum effects cannot be neglected. 
These effects may thus be seen as a kind of smile of a Cheshire quantum gravity cat, left over
when the actual classical curvature can be ignored.

\subsection{Related mathematical issues}

In ordinary special relativity there is no such limitation to how far apart two clocks may be and be synchronized, and hence no limit to how large the coordinate
frame used by an inertial observer may be.  Because of this, the Poincare transformations form a Lie group.  Consider boosts between observers at
rest and moving with a velocity $v$ made at the two events
$\cal E$ and ${\cal F}_1$ and call them $B_{\cal E}(v)$ and $B_{{\cal F}_1}(v)$.  If $T$ is the translation that takes the event to  ${\cal F}_1$ to $\cal E$ we have
\f
B_{\cal E}(v) =   T \cdot     B_{{\cal F}_1}(v) \cdot T^{-1}
\label{sdp}
\ff

In other words, using the fact that Poincare group is a Lie group that contains translations and boosts, one can extend the definition of an inertial frame to coordinatize all of Minkowski spacetime without ambiguity.  This expresses the fact that Einstein's procedure for synchronizing moving clocks can be extended to clocks arbitrarily far away from a reference clock.

However, as we have just seen here that the situation cannot be so simple  in $\kappa$-Minkowski spacetime. 
This raises the question of whether the $\kappa$-Poincare algebra can be exponentiated to a group, and what the structure of that group is.  This is
a question that has been investigated mathematically and it has been shown that, while the $\kappa$-Poincare algebra has subalgebras which are isomorphic to the ordinary Lorentz algebra, that do exponentiate to the Lorentz group, the whole of the $\kappa$-Poincare algebra does not exponentiate into an ordinary Lie group\cite{majid,JM2}.
This question is outside the scope of this paper, but the results of this paper underlie its importance. 

\subsection{A perspective on these results}

What we can do is comment on the meaning of these results in the context of the general problem of the construction of a quantum theory of spacetime.    Recall Mach's principle, or more generally, the notion of relational spacetime, which asserts that the spacetime geometry reflects the relationships amongst physical particles whose history actually defines the spacetime.  This idea is implicit in special relativity; indeed 
Einstein defined events in special relativity by the coincidence of worldlines representing the histories of physical particles. In special relativity however the lorentz transformation of an event does not depend on the properties of a particle traveling on the worldlines, as a result one has an unambiguous description of spacetime independent of what travels through it.  This allows one to abstract the spacetime geometry away from the detailed history of the particles whose interactions actually define it.   This abstraction results in the geometry of Minkowski spacetime.  

However this abstraction leads to a conceptual tension,  arising from a conflict between the interpretation of Minkowski spacetime  in special and general relativity.  Einstein tells us that in special relativity the points of Minkowski spacetime are operationally defined as arising from the coincidence of physical particles.  But in general relativity, Minkowski spacetime is a vacuum solution, corresponding to a limit in which all matter has been removed from spacetime.  The usual resolution of this tension is to regard special relativity only as an approximation to general relativity, which more fully realizes 
the principle that spacetime is relational.

One can ask whether this tension could also arise in the context of a quantum gravity theory.  It is then interesting to note that  it not possible to completely abstract the spacetime geometry from the motions and interactions of particles in $\kappa$-Minkowski spacetime. As we have seen, the lorentz transformations (\ref{xtrans},\ref{ttrans}) and translations (\ref{translate})  of the coordinates of an event depend on the energy and momentum of the particles whose intersection defines the event.    

Indeed, hidden in the form of  the energy and momentum dependence of (\ref{xtrans},\ref{ttrans}) and  (\ref{translate}) is a question, what energy and momenta are involved here?  From the context in which these formulas are derived\cite{JM1,JM2} the answer is clear, these transformations apply to the phase space of a free relativistic particle, whose coordinates and momenta and energy are always defined.  But this answer raises a further question: can we abstract from the phase space description of a free particle  to a universal description of a spacetime geometry that is independent of the particular particles
traveling through it?  To do this we would need a formula for lorentz transformations of events that do not depend on particular properties of the particles whose motion and interaction defines that event. Without this there can be no abstraction to an empty spacetime geometry from the phase space description of a relativistic particle.  

Of course such formula are available, by taking the formal limit in which $l_p E/  \hbar \rightarrow 0$ one returns to the description in special relativity.  However, looking at (\ref{xtrans},\ref{ttrans}) and  (\ref{translate}) we see that the relevant limit is really of the form
$l_p x E / \hbar  \rightarrow 0$.  This is non-uniform in space, hence for a given scale of energy it can only be taken within an horizon in which
$|x|$ is not too large.  This is how the limitations of the extent of inertial coordinates we have just discussed arises.  It is a limit to the region within in which one can abstract
from the phase space a particle to get an independent notion of spacetime geometry, in which events are defined without regard to the particles whose interactions define them.  If we don't take the limit $l_p x E/ \hbar  \rightarrow 0$ then there is no notion of the coordinates of an event that does not carry with it, coded into its transformation properties, a record of the energy and momentum of the particles whose interaction defined it.  That is, there simply are no events without particles.  

Hence, the ambiguities in coordinates of events we have discussed here can be seen in a new light, they indicate that in quantum gravity we have to take seriously the notion that spacetimes are constructed to represent relations amongst physical events involving physical particles.  There is in general no
abstraction which yields an empty spacetime geometry, in which events are defined without reference to the particles that operationally defined them.  

Mathematically, these ambiguities reflect the uncertainty relation (\ref{nup}) arising from the commutation relations (\ref{[x,t]}).  These formulas are also indications of the basic lesson that in any experimental situation in which quantities of the order of $l_p E x /\hbar$ cannot be neglected there is a fundamental limit to the use of the notion of a spacetime geometry abstracted from the network of causal interactions amongst physical degrees of freedom.  

This makes sense of the fact that in $DSR$ theories the primary arena for physics is the momentum space, whereby spacetime is a derived quantity.
The limitations we have been discussing arise when one attempts to abstract a description in spacetime alone, the lesson is there is a limit to the meaning of the pure spacetime description, absent the information from momentum space. 

In this light we can also mention, but not address, another issue that has both fundamental conceptual and mathematical aspects.  The transformation formulas
(\ref{xtrans},\ref{ttrans}) and  (\ref{translate}) that we are discussing arise in the phase space description of a single free relativistic particle\cite{JM1,JM2}.  However, in our discussion we make use of Einstein's principle that the definition of spacetime events arises from the interactions of particles. To further 
investigate quantum geometry we must then make use of a mathematical description which involves many particles and incorporates interactions.
In this light we can note that in $\kappa$-Minkowski spacetime multiparticle states are constructed by a non-trivial procedure involving the co-product, which extends the usual direct product that is the basis of the construction of Fock space\cite{majid,kappafields,nonlocality}. Elucidation of this construction is
needed both to incorporate interactions into $DSR$ theories and to understand the physical interpretation of quantum and non-commuting spacetime geometry.

\section{Conclusions}

In this paper we have studied the problem of non-locality in the emission and detection of gamma
ray bursts, in a well studied formulation of $DSR$, which is $\kappa$-Minkowski spacetime. We find that there is no macroscopic non-locality at
the detector in either the lab frame or a frame moving with respect to it.  We do find that there are ambiguities in the assignment of coordinates
to events distant from the origin of inertial frames, which can become macroscopic when the events are at cosmological distances from the 
observer.  We have proposed that these are not real physical non-localities, but just ambiguities in the procedure by which inertial observers
assign coordinates to distant events. 

These results disagree with those of \cite{sabine-paradox}; the reasons for the disagreement are discussed in the Appendix.   

In closing, we note that an  issue of a macroscopic non-locality at the detector, seen by an observer located there, would be very different from the issue of a possible macroscopic non-locality that arises only at a cosmological distance from the observer.  To establish that a cosmologically distant event is afflicted by a physical non-locality, rather than a coordinate artifact, one would have to be assured that the coordinates defined by a process of synchronization with the clock of a local observer are unambiguously defined arbitrarily far from that clock.   Here we have examined whether this is the case for 
$\kappa$-Minkowski spacetime, and found that there are ambiguities in  the sychronization procedure coming from either the energy dependence of the speed of light or the commutation relations (\ref{[x,t]}). 

The claim that the problem is a coordinate artifact and not a physical non-locality is also strengthened 
by the fact that there is in the expression for $\Delta t_{{\cal E}^\prime}$ in (\ref{effect1},\ref{effect2}) a factor of $v$, which is the velocity of the moving frame we are transforming to.  Hence 
which event is earlier or later depends on the relative motion of two observers!  This makes no sense for a real physical effect-but is easily understood if the expression
refers to an ambiguity in the procedure to synchronize clocks and construct coordinates.

Hence we encounter a new situation in realizing the relativity of inertial frames in a non-commutative spacetime, which is that a feature of special relativity fails to hold here.
In special relativity the synchronization of clocks can be extended arbitrarily far from the clock which anchors the procedure.  This of course breaks down when there is curvature.  Here we see the process by which an inertial observer assigns coordinates to distant events is breaking down even in the absence of the gravitational field. This is due either to effects of the non-commutativity of spacetime (or, equivalently, the curvature of momentum space), or
to an energy dependence of the speed of light.   This should not be surprising, to the contrary it would have been strange if the principle of the relativity of inertial frames can be extended to the non-commutative case without loosing some feature of the physics.  Giving up the presumption that the
procedure of clock synchronization can be carried out to cosmological distances is not a lot to give up.  Indeed, in the real world, the procedure breaks down for curvature effects long before the issues discussed here are encountered.  

This is then a situation like curvature, or the obstructions in the construction of coordinate patches in complex manifolds, in which a procedure for
assigning meaningful coordinates breaks down globally.  This raises the question of whether there is an elegant global description as we do find
in curved geometries and complex manifolds, perhaps inherent in the mathematical structure of the $\kappa$-Poincare algebra\cite{majid}.  If so, this may affect other open problems such as the meaning of velocity of distant
particles in $\kappa$-Minkowski spacetime\cite{velocity-first,Tezuka}, the problem of unitarity\cite{notunitary}, as well as the issue of constructing sensible
interacting quantum field theories in $\kappa$-Minkowski spacetime{kappafields}. 

These remain crucial  issues for further work.

\section*{ACKNOWLEDGEMENTS}

I am grateful to Sabine Hossenfelder for many discussions and emails seeking to clarify these issues.  I am also grateful to Michele Arzano, Laurent Freidel,  Jurek Kowalski-Glikman, Joao Magueijo, Shahn Majid, Seth Major, Chanda Prescod-Weinstein and especially Giovanni Amelino-Camelia for related discussions and correspondence.  I am also especially grateful to Michele Arzano and Jurek Kowalski-Glikman for showing me a draft of the relevant chapters of their book\cite{JM2} ahead of publication.  
Research at Perimeter Institute for Theoretical Physics is supported in part by the Government of Canada through NSERC and by the Province of
Ontario through MRI.

\appendix

\section{Relationship to other approaches}

In this appendix I would like to contrast the results we have found to hold for $\kappa$-Minkowski spacetime with those found
for other approaches.  

First, there is an analysis of a different formulation of $DSR$, based directly on deforming the action of the generators
of the Poincare algebra\cite{whataboutbob}, which obtains results similar to those found here:  There is no macroscopic non-locality in the detection events, seen by an observer near the detector, but there are issues of coordinate ambiguities for the emission event.

On the other hand, \cite{sabine-paradox} finds that Bob's satellite frame sees macroscopic non-locality at the detector.  It is important to understand
the reason for this disagreement.   We first look at the argument of \cite{sabine-paradox}, then we contrast its assumptions with those
made here.

\subsection{The claim that $DSR$ implies macroscopic non-locality}

The conclusions of reference \cite{sabine-paradox}  are arrived at by reasoning directly from several assumptions to a conclusion about how the experiment would look in Bob's frame of reference.  The first three of these assumptions can be taken to be

\begin{itemize}

\item{}{\bf A}) Both observers can describe the photons' trajectories accurately in terms of worldlines in Minkowski spacetime.

\item{} {\bf B}) Both observers assume that the same equation (\ref{c(E)}) holds in their frame's coordinates.

\item{} {\bf C}) Both observers assume that the two photons left the gamma ray burst from the same point at the same time, ie at a single value of the
spacetime coordinates in each frame\footnote{It should be emphasized that $\bf C$ is not explicitly stated in \cite{sabine-paradox}, however it is the most natural
interpretation of Figures 1 and 2 of that paper. (Note the worldlines corresponding to the two photons crossing in both diagrams in the lower left hand corner.)  $\bf C$  is equivalent, within the framework of \cite{sabine-paradox}, to 
 the statement that the formula for the time delay $\tau$ (\ref{timedelay}) should hold
in all frames of reference.   It is also equivalent to assuming that the fixed point of the boost from the lab to the satellite frame is at the emission point, at least
in the framework discussed here.}. 

\end{itemize}

Given these assumptions, simple algebra leads to the conclusion that Bob does not see the electron and the second photon to pass through the detector at the same time.  Indeed, with $L$ at cosmological distances and $\Delta E \approx 10 Gev$, as might describe a real event detected by Fermi, they miss by about $10^{-5}$ of a second, which given that the second photon is going almost the speed of light amounts to a very large miss in terms of spatial distance.

From this \cite{sabine-paradox} concludes that either there is an inconsistency in the theory (because it is claimed that that the three assumptions ${\bf A, B}$ and ${\bf C}$ follow from the relativity of inertial frames) or Bob must deduce that there are in physics highly non-local interactions which allow the second photon to scatter from the electron in spite of being separated by $10^{-5}$ light seconds.  

\cite{sabine-paradox} then argues that there are good experimental bounds on non-locality, which applied to this situation restrict $\alpha < 10^{-23}$.  This part of the argument rests on 
a fourth assumption

\begin{itemize}

\item{}{\bf D}  The kind of non-local interactions produced by lorentz transformations in DSR theories are sufficiently generic that existing bounds on ranges of 
interactions from experiments that do not involve moving frames observing light that has traveled for billions of light years can be used to bound $\alpha$
in this experiment.  

\end{itemize}

\subsection{Comparison of the results of the different approaches}

We can now compare the results of   \cite{sabine-paradox} with those found here for $\kappa$-Minkowski spacetime.  

The case most relevant for the assumptions of \cite{sabine-paradox} is that of the commuting coordinates, as these lead to an energy dependent
speed of light.  There are two differences.  In \cite{sabine-paradox} there is macroscopic non-locality at the detector seen in Bob's satellite frame,
whereas in $\kappa$-Minkowski spacetime the non-locality seen by Bob is much smaller, of order $t_p^2$.  

The second difference is that in \cite{sabine-paradox} the emission event is fixed by assumption $\bf C$ to have a single coordinate in all reference frames, whereas in $\kappa$-Minkowski spacetime   Bob's description the emission event has now two time coordinates which differ
by  $\Delta t_{{\cal E}^\prime} = t_p L \Delta E$.  This means that assumption $\bf C$  fails.  The apparent splitting of the event is
macroscopic, but the different is that the issue of ambiguity or non-locality has to do with an event a cosmological distance from the observer, rather than at the observer.  

It is interesting to note that $\bf C$ fails both on the assumption that the particles travel on worldlines defined by the commuting and the non-commuting 
coordinates.  This remains true whether the emission of both photons is defined to happen in Alice's frame at constant $t$ or
constant $T$.  

The crux of the disagreement between the conclusions of \cite{sabine-paradox} and the present calculations is what event is held fixed in the boost made from Alice to Bob's frame of reference.  Every pure boost has a unique fixed point whose coordinates are the same in the coordinate system of both frames.  As we have argued, given that Alice's lab frame and Bob's satellite frame coincide at a unique event when they pass each other at the event  ${\cal F}_1^a= (0,0)$, it is natural to define Bob's coordinates so that this event is the fixed point whose coordinates are left unchanged by the boost.  

In \cite{sabine-paradox} it is assumed instead that the transformation from Alice to Bob's frame has its fixed point at the emission event. 
This is justified in \cite{sabine-paradox} by the condition that the equation
for the time delay $\tau = l_p \Delta E L/ \hbar $  (\ref{timedelay}) holds in all reference
frames. This implies that $\bf C$ holds. 

To resolve this disagreement, note that the issue is not about how to do a lorentz transformation in a $DSR$ theory, it is about which lorentz transformation gives the correct
coordinates for observations measured by the Bob's satellite frame.

It is of course possible to define the transformation from Alice's lab frame to a frame whose origin of coordinates is at the emission point; this is a combination of a translation from the detection event to the emission event,  followed by a pure boost around that point.  This gives the description of the experiment in terms of coordinates that would be used by a moving observer whose origin coincides with the emission event.  If one did that then the macroscopic non-locality is an issue at the detector-because the non-locality is always macroscopic at a coordinate z such that $l_p z E/\hbar$ is measurable, and this
means $z$ must be a cosmological distance.  This is how
the result of \cite{sabine-paradox} that there is macroscopic non-locality at the detector is arrived at. 

The point is that this is not the right calculation to do to arrive at the measurements made by Bob's satellite frame, whose origin is at the
event $(0,0)$.   This is a matter of definition, it is what is required by the
procedure which defines the coordinates of the inertial frame.  

While the definition should suffice, if one needs additional support for this choice of the fixed point, one can also argue that no other event is unique in this way.  Bob is a general purpose observer, he can observe photons from many different distant events.  If we let the definition of Bob's coordinates depend on one particular distant event, what picks out that event? 

Another problem with making the fixed point depend on the emission of a gamma-ray burst is that it is impossible to carry out in practice.  Imagine the actual case of the Fermi satellite synchronizing its time and space coordinates with measurements made by a gamma ray detector on the ground\footnote{Note that in \cite{sabine-paradox} the detector is on the ground and the satellite is just observering events in the detector.  This is the reverse of the usual situation, but of course doesn't affect the arguments.}.  To synchronize its coordinates so that an emission event was the fixed point of the boost, the satellite would need a measurement of the redshift of that gamma ray burst.  This is not always available.  Does this mean that the Fermi satellite cannot synchronize its description of a gamma ray burst with those of a detector on the Earth if it happens, as is often the case, that no redshift is obtained for that gamma ray burst?

Furthermore, one cannot maintain $\bf C$ and the observer independence of the form
of the time delay (\ref{timedelay}) for more than a single gamma ray burst.  
Suppose the detector on the ground observed five gamma ray bursts, to which should the satellite synchronize its measurements?   

For these reasons it seems best to use the unique point of intersection of the origins of the two reference frames as the fixed point of the boost between the two frames and give up 
assumptions $\bf C$ and the frame independence of (\ref{timedelay})\footnote{For the contrary case argument \cite{sabine-paradox,sabine-reply}.}.  
One cannot maintain these assumptions for more than one of the many gamma ray bursts that are detected, so why not give them up for all in favor
of carrying over to $DSR$ the usual well defined perscription used in special relativity for defining the boost to a moving frame of reference?  
 
One might reply that it cannot matter where the fixed point is, because Minkowksi spacetime is homogeneous.  Indeed, so is $\kappa$-Minkowksi spacetime,
as shown by the presence in its symmetry algebra of deformed generators of translations.  This homogeneity implies one can put the clock which defines
the origin of an inertial reference system anywhere in spacetime.  But it does not imply that each observer can unambiguously assign coordinates to a fixed
accuracy to events arbitrarily distant from them.  

\subsection{State of the issue of macroscopic non-locality}

Reference \cite{sabine-paradox} made a very strong claim, that there is  " ... a conflict with experiment to very 
high precision\cite{sabine-paradox}...If DSR was indeed the origin of time-delays of highly energetic photons from GRBs, then it would also  lead to macroscopic effects we would long have observed. 
Consequently, DSR cannot be cause of observable effects in GRB spectra\cite{sabine-paradox}."  

A claim that an experimental bound exists on a parameter of a new physical theory requires a solid case. It can be weakened or undermined if
there are unresolved issues of physical interpretation, or there exist versions or interpretations of the theory in question to which the argument does not apply.   For this reason
there have been a number of papers discussing the claim\cite{whataboutbob,me-firstresponse,sabine-reply,JM1,seth-DSR10}.

I would like to  summarize my understanding of the present status of this issue:  

\begin{enumerate}

\item{}  The claim that there is a bound from existing experiments on the parameters of $DSR$ theories rests on the claim that the satellite frame would see macroscopic non-local effects at the detector.  

\item{}  What the satellite frame sees can be found by making a pure lorentz boost from the lab frame. ÊIn the framework of  $\kappa$-Poincare, this Lorentz boost is completely specified by the condition that the origins of the 
lab and satellite frames coincide, and coincide with the event at which their two world lines intersect, as it would be in special relativity.  
The correct lorentz transformation in this case is the pure boost that fixes that unique event where worldlines of the two detectors intersect.

\item{} ÊWhen this Lorentz transformation is computed explicitly in $\kappa$-Poincare there is no macroscopic non-locality in the satellite frame's description of the detector.   ÊThis is the case in both the commuting and non-commuting spacetime coordinates.  Computations
in other formulations of $DSR$ agree \cite{whataboutbob}.  

\item{}The potential non-localities that are produced by Lorentz boosts at the detector are on distance scales of the order of $v L E^2/c E_p^2 $, (\ref{realresult}).
This disagrees with the prediction of \cite{sabine-paradox},  by a factor of 
$E/E_p \approx 10^{-18}$.  It is much smaller than the Compton wavelength of an electron in the detector and therefor unobservable.  

\item{} Therefore the  claim in \cite{sabine-paradox}, of  an experimental  bound on the $\alpha$ in (\ref{c(E)}), does not apply to $DSR$ theories in general, because the prediction on which it is based, of a macroscopic non-locality at the detector, does not appear in at least one formulation of $DSR$.   
Hence, the experiments testing the consequences of (\ref{c(E)}) need to be done, and the results of current observations by Fermi and other astrophysical detectors cannot be predicted in advance based on the knowledge we have now as to the fate of lorentz invariance at the Planck scale.  

\item{} The reason the  calculation of  \cite{sabine-paradox} disagrees with that done here  is that it makes the assumption 
that the fixed point of the lorentz transformation from the detector to the satellite frome is at the emission event.    This is equivalent to 
$\bf C$, which is that there is no ambiguity in the coordinates of the emission event.  
 But when one defines the fixed point of the boost to the satellite frame to be the unique event where the worldlines of the two observers coincide, at the detector, explicit calculations  show that there are ambiguities in the definition of the coordinates of the emission event.  Hence, $\bf C$ is not satisfied.  This in an example of a general phenomena in which lorentz boosts can produce energy dependent ambiguities in the coordinates of
events very distant from the origin of coordinates of an inertial frame. 

\item{}We can mention also that assumptions $\bf A$ and $\bf D$ are critiqued in \cite{me-firstresponse} and discussed further in \cite{sabine-reply}.  

\end{enumerate}

It is then certainly necessary to investigate the status of the ambiguities in the coordinates of distant events in boosted frames. ÊThis was the second aim of this paper.  There are two points of view:

{\bf a)} ÊThey indicate a real physical non-locality.

{\bf b)} ÊThey are the result of a coordinate ambiguity, because of an ambiguity in applying Einstein's procedure to synchronize moving clocks far from the origin of a frame.

It is clear that if {\bf a)} is correct then the problem of a physical macroscopic non-locality is just shifted to the event where the photons are emitted. However if {\bf b)} is correct than there is no macroscopic non-locality at either the gamma ray bust or the detector and no argument from non-locality to an experimental bound.   The aim of this paper has been to propose the case for {\bf b}.  

Hence, even if we find results which disagree with those of \cite{sabine-paradox}, 
the challenges posed by that argument are  highly non-trivial and are already leading to a deeper understanding of the physics of $DSR$. 

\section*{NOTE ADDED August 9,  2010}

In a preprint, Hossenfelder has offered a critique of this paper\cite{sabine-critique}.  To address the key point, \cite{sabine-critique} appears to agree that there are ambiguities in synchronization of distant clocks but disagrees that they can explain the apparent non-localities produced by lorentz transformations.  The argument in \cite{sabine-critique} however fails to address the claims of this paper because it is based on the weaker limit, (\ref{horizon2}) and (\ref{nup2}) and ignores the stronger limit (\ref{horizon1}) and (\ref{horizon3}) which takes into account the energy of the photon observed at the distant event.  As (\ref{horizon2}) and (\ref{nup2}) do not involve the energy of the photon detected they obviously cannot account for apparent non-localities that do depend on the photon's energy; it doesn't need an elaborate development to establish this.  But as we argued above, the limit on synchronization must depend on the energy of the energetic photon if the coordinate system is not to be thrown out of synchronization by the distant clock's absorption of the energetic photon.  This leads to (\ref{horizon1}) and (\ref{horizon3}), and when this is taken into account, one concludes that the coordinate ambiguities produced by using exchanges of light signals to define coordinates does explain the apparent non-localities produced by boosting to moving reference frames. 

Several of the other points raised in \cite{sabine-critique} are already addressed in the above appendix.
While the discussion is complicated by a confusing and non-standard use of the term ``fixed point", it remains the case that the large apparent non-localities that result from a lorentz boost always occur far from the origin which is the fixed point (with the usual meaning) of that lorentz boost.  With regard to the discussion of the assumptions of this paper, it is simplest just to say that I am proposing to apply to DSR theories exactly the same operational procedure for the construction of the coordinates of an inertial observer that Einstein used in his original papers on relativity.  

Finally, it is true that no proposal has been yet made for how to construct a global mathematical framework that patches together the limited local non-commutative inertial frames, so we can agree that more work needs to be done to develop the proposal of this paper.

\end{document}